\documentclass{iau} 
\usepackage{graphicx}
\usepackage{hyperref}
\usepackage{multicol}

\title[JD 01.Gamma-ray Unidentified sources] 
{Multiwavelength study of potential blazar candidates among \textit{Fermi}-LAT unidentified gamma-ray sources }

\author[Jean Damasc\`{e}ne Mbarubucyeye, Felicia Krau\ss ~\& Pheneas Nkundabakura]   
{Jean Damasc\`{e}ne Mbarubucyeye$^1$, Felicia Krau\ss$^2$,
 \and Pheneas Nkundabakura$^3$}

\affiliation{$^1$Deutsches Elektronen-Synchrotron (DESY),
Platanenallee 6, 15738 Zeuthen, Germany, \\ email: {\tt mbjdamas@gmail.com} \\[\affilskip]
$^2$Department of Astronomy \& Astrophysics,
  Pennsylvania State University,
  University Park, PA 16801, USA \\email: {\tt Felicia.Krauss@psu.edu} \\[\affilskip]

$^3$University of Rwanda, College of Education,
P.O.  Box 5039, 
Kigali, Rwanda  \\email: {\tt nkundapheneas@yahoo.fr}}

\pubyear{2019}
\volume{356}  
\jname{Nuclear Activity in Galaxies Across Cosmic Time}
\editors{M. Povi\'c, P. Marziani, J. Masegosa, H. Netzer,\\ S. H. Negu, \&
	S. B. Tessema, eds.}
\begin{document}

\maketitle

\begin{abstract}
Studying unidentified $\gamma$-ray sources is important as they may hide new discoveries. We conducted a multiwavelength analysis of 13 unidentified Fermi-LAT sources in the 3FGL catalogue that have no known counterparts (Unidentified Gamma-ray Sources,
UnIDs). The sample was selected for sources that have a single radio and X-ray candidate counterpart in their uncertainty ellipses. The purpose of this study is to find a possible  blazar signature and to model the Spectral Energy Distribution (SED) of the selected sources using an empirical log parabolic model. The results show that the synchrotron emission of all sources peaks in the infrared (IR) band and that the high-energy emission peaks in MeV to GeV bands. The SEDs of sources in our sample are all blazar like. In addition, the peak position of the sample reveals that 6 sources (46.2\%) are Low Synchrotron
Peaked (LSP) blazars, 4 (30.8\%) of them are High Synchrotron Peaked (HSP) blazars, while 3 of them (23.0\%) are Intermediate Synchrotron Peaked (ISP) blazars.
\end{abstract}

\firstsection 
\section{Introduction}

The Large Area Telescope (LAT) on board the \textit{Fermi} Gamma-ray Space Telescope detects high-energy photons between 20\,MeV and 300\,GeV  (\cite[Atwood et al. 2009]{Atwood2009}). In the 3FGL catalog among 3033 sources detected, 33\% are still unassociated to any counterparts (\cite[Acero et al. 2015]{Acero2015}) while in the latest catalog (4FGL) more than 38\% remain unassociated (\cite[The Fermi-LAT collaboration 2019]{Collabo2019}). Among 2023 sources that have been already identified in 3FGL, 1100 are Active Galactic Nuclei (AGN) of blazar type (~57\%).
Blazars are a subclass of AGN with their relativistic jets pointing at Earth. The two main classes of blazars according to their optical properties are: (1) Flat Spectrum Radio Quasars (FSRQs) characterised by the strong emission lines and (2) BL Lacs
characterised by weak or no emission lines at all. In addition, blazars are classified according to the position of their peak frequency (\cite[Massaro et al.2004]{Massaro2004}): Low Synchrotron-Peaked (LSP) objects with $\rm{\nu_{p}^{synch}\leq 10^{14}\,Hz}$, Intermediate Synchrotron-Peaked (ISP) objects with $\rm{10^{14}\,Hz\leq \nu_{p}^{synch}\leq 10^{15}\,Hz}$, and High Synchrotron-Peaked (HSP) objects with $\rm{\nu_{p}^{synch}\geq 10^{15}\,Hz}$.
In this research, we determined the type of emissions from a sample of 13 blazar candidates and classified them according to the shape of their SEDs.

\section{Sample selection}
\label{sect.2}
 \noindent The 3FGL catalogue includes 1010 unidentified sources~(\cite[Acero et al. 2015]{Acero2015}). The release of the 3FGL catalogue enabled the \textit{Swift}-XRT survey of the \textit{Fermi}-LAT unassociated sources with the purpose of performing their follow-up observations in an attempt to find their potential X-ray counterparts. 
 Two common features that all blazars share are that they are radio emitters (radio-loud objects) and that they are highly variable. \cite[Yang \& Fan (2005)]{Yang2015} showed that there is a strong correlation between radio and $\rm{\gamma}$-ray emission in blazars when they are in a their flaring high state. However, the most recent studies are continuously debating about the correlation between radio to $\rm{\gamma}$-ray bands (\cite[Massaro et al. 2017]{Massaro2017}; \cite[Bruni et al. 2018]{Bruni2018}). The selected sample takes advantage of the fact that most of the studied blazars exhibit strong radio-X-ray emission (\cite[Padovani et al. 2007]{Padovani2007}). In addition, most of the known $\gamma$-ray bright AGN are above a Galactic latitude of $|b|> 10$ degrees, due to possible source confusion as well as higher diffuse background in the Galactic plane (\cite[Ackermann et al., 2015]{Ackermann2015}). A sample of unassociated sources was selected based on the following criteria:
\begin{enumerate}
 \item Only a single XRT detection lies in the uncertainty ellipse of the 3$\sigma$ of the LAT unassociated source.
 \item Only a single radio source lies in the uncertainty box of \textit{Fermi}-LAT uncertainty region which is coincident with the XRT detection.
 \item Only sources with $|b| >10$  degrees in order to limit for any source confusion in the Galactic plane.
\end{enumerate}
Applying all cuts to the population of unassociated sources listed in the 3FGL and observed by \textit{Swift}, we isolated a sample of 13 unassociated sources assumed to be potential blazar candidates.

\section{Data analysis}
\label{sect.3}
\noindent The \textsl{Swift}-XRT/UVOT data available were considered. We used standard extraction methods to produce \textsl{Swift}/XRT and \textsl{Swift}/UVOT spectra. \par

\noindent The \textsl{Fermi}-LAT spectral data point were extracted using Fermi Science tools (v10r0p5 released on June 24, 2015). The analysis takes into account the source region and the region of interest (ROI). We selected events within the energy range $\rm{100\,MeV-300\,GeV}$, a maximum zenith angle of 90 degrees, a ROI of 10 degrees. \par

\noindent A complete multiwavelength SED was obtained by supplementing the XRT/UVOT and \textit{Fermi}-LAT data with other data from across the whole electromagnetic spectrum. These data were obtained using SEDbuilder tool\footnote{\url{http://www.asdc.asi.it/}}. \par

\noindent In order to identify the type of emission from the source, we apply an empirical model on the data. The SED shows that the log-parabolic model fits the data well for blazars, in different energy bands (\cite[Massaro et al. 2004]{Massaro2004}; \cite[Tramacere et al. 2009]{Tramacere2009}). The model used includes two log parabolas, one fitting the low-energy spectrum (synchrotron emission) and another one fitting the high-energy spectrum. In addition to these two log-parabolas, three other components are added: absorption, extinction, and a blackbody. 

\section{Results and discussions}
\label{sect.4}
\noindent The SEDs resulted from the fitting and reported herein show the two typical bump known as the main signature of blazars. The low-energy bump (peaking between IR and optical bands) is interpreted as synchrotron emission from highly accelerated relativistic electrons and the high-energy bump (peaking at X-ray and $\rm{\gamma}$-ray energy bands) is related to high-energy emission, which could be either leptonic (Inverse Compton) or hadronic in nature.
Based on the position of the $\nu{}^\mathrm{sync}_\mathrm{peak}$ (\cite[Abdo et al., 2010]{Abdo2010}), we found that 6 ($\sim 46.2\%$) sources of our sample are LSP objects, 3 ($23.0\%$) sources are ISP objects and 4 (30.8\%) of them are HSP objects.  
Figure 1 shows an example of the SED for 3FGL J1220.1$-$3715 where data are fitted by the model. The lower panel represents residuals.
\begin{figure}
	\centering
\includegraphics[width=0.6\columnwidth]{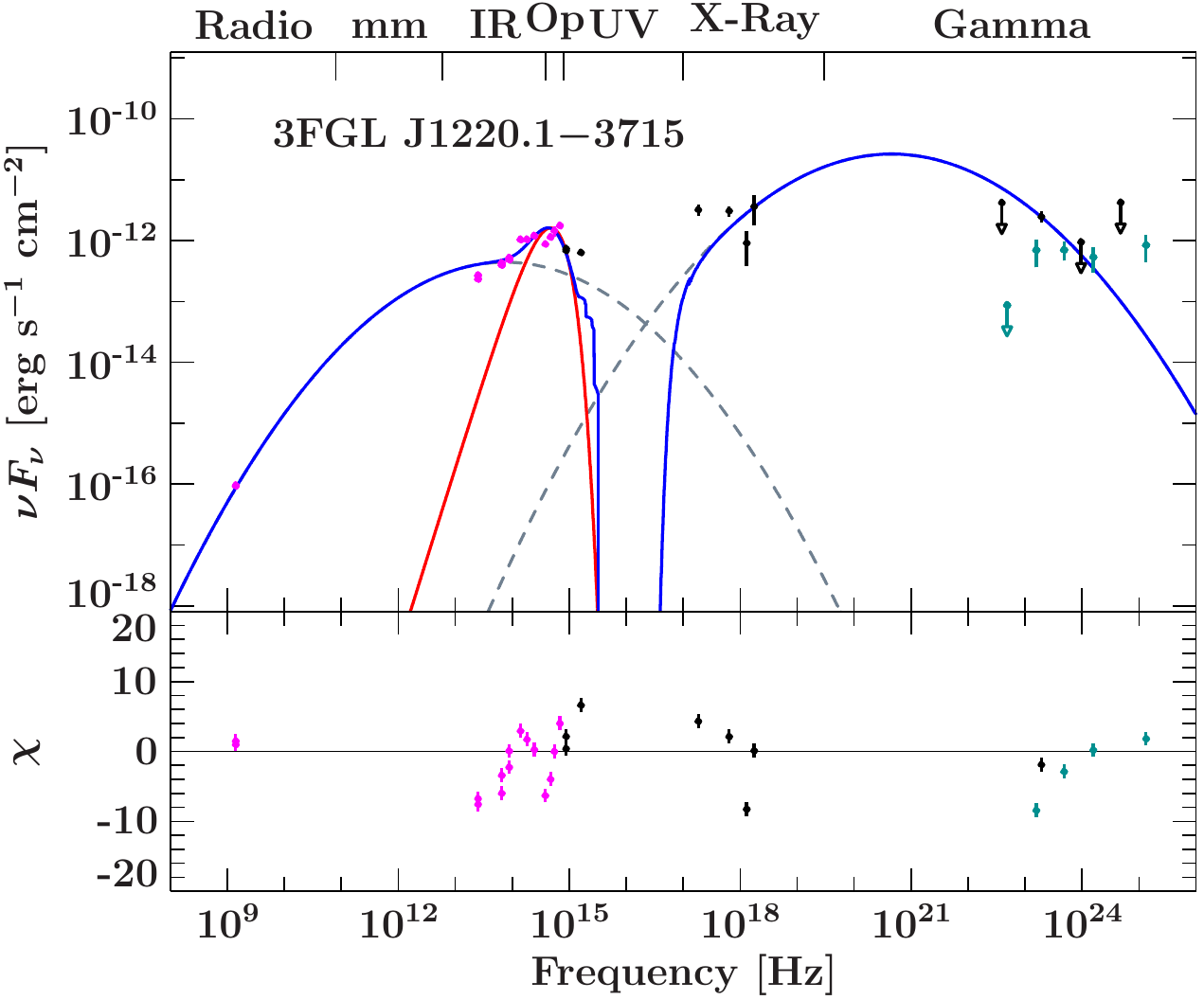}  
\caption{A sample of modeled SED of our sample with a log parabolic model. The model includes two absorbed logarithmic parabolas shown in blue, a blackbody in red dashed, and the total unabsorbed model shown in black. The extracted data are black colored while 3FGL data are in cyan and non-simultaneous archival data are in magenta.}
\label{1}
\end{figure}
\noindent For most of the cases, our classification is in agreement with \cite[Parkinson et al. (2016)]{Parkinson2016}, \cite[Lefaucheur \& Pita (2017)]{Lefaucheur2017} and \cite[Salvetti et al. (2017)]{Salvetti2017}, who classified the sources of our sample as AGN and blazars. All three works use Machine Learning techniques. We finally used the WISE IR colour-colour plot taking into account sources that have WISE matching sources. Only 9 sources have matching counterparts in WISE catalogue. As a results we found that only 7 sources lies in the WISE blazar region which emphasize their candidacy to be blazars.
\footnotesize

\end{document}